\documentclass[journal]{IEEEtran}
\usepackage{cite}
\usepackage{amsmath,amssymb,amsfonts}
\usepackage{algorithmic}
\usepackage{graphicx}
\usepackage{caption}
\usepackage{subfig}
\usepackage{textcomp}
\usepackage{xcolor}
\usepackage{color, soul}
\usepackage{enumitem}
\usepackage{import}
\usepackage[ruled,vlined]{algorithm2e}
\usepackage{setspace}
\usepackage{hyperref}
\usepackage{tabularx}
\usepackage{comment}
\newcommand\myfontsize{\fontsize{23pt}{28pt}\selectfont}

\def\BibTeX{{\rm B\kern-.05em{\sc i\kern-.025em b}\kern-.08em
    T\kern-.1667em\lower.7ex\hbox{E}\kern-.125emX}}
\begin{document}
 
% \title{Overview of Distributed Energy Resource Cybersecurity: Vulnerabilities, Attacks, Impacts, and Mitigations}

\title{\myfontsize{Distributed Energy Resources Cybersecurity Outlook: Vulnerabilities, Attacks, Impacts, and Mitigations}

\thanks{\rule{4cm}{0.5pt}\\
% received 20 May 2022; revised 17 December 2022 and 16 June 2023; accepted 9 August 2023.
Manuscript accepted in IEEE Systems Journal.\\
Ioannis Zografopoulos is with the King Abdullah University of Science and Technology, Thuwal 23955, Saudi Arabia (e-mail: zografop@gmail.com). \\
Nikos D. Hatziargyriou is with the National Technical University of Athens, 10682 Athens, Greece, and the University of Vaasa, 65200 Vaasa, Finland. \\
Charalambos Konstantinou is with the King Abdullah University of Science and Technology, Thuwal 23955, Saudi Arabia.}}

\author{Ioannis Zografopoulos,~\IEEEmembership{Graduate Student Member,~IEEE},  {Nikos D. Hatziargyriou},~\IEEEmembership{Life Fellow, IEEE}, {Charalambos~Konstantinou,~\IEEEmembership{Senior Member,~IEEE}} 
}

\IEEEaftertitletext{\vspace{-2.5\baselineskip}}

\maketitle

\begin{abstract}
\textcolor{black}{The digitization and decentralization of the electric power grid are key thrusts for an economically and environmentally sustainable future. Towards this goal, distributed energy resources (DER), including rooftop solar panels, battery storage, electric vehicles, etc., are becoming ubiquitous in power systems. Power utilities benefit from DERs as they minimize operational costs; at the same time, DERs grant users and aggregators control over the power they produce and consume. DERs are interconnected, interoperable, and support remotely controllable features, thus, their cybersecurity is of cardinal importance. DER communication dependencies and the diversity of DER architectures widen the threat surface and aggravate the cybersecurity posture of power systems. In this work, we focus on security oversights that reside in the cyber and physical layers of DERs and can jeopardize grid operations. Existing works have underlined the impact of cyberattacks targeting DER assets, however, they either focus on specific system components (e.g., communication protocols), do not consider the mission-critical objectives of DERs, or neglect the adversarial perspective (e.g., adversary/attack models) altogether. To address these omissions, we comprehensively analyze adversarial capabilities and objectives when manipulating DER assets, and then present how protocol and device-level vulnerabilities can materialize into cyberattacks impacting power system operations. Finally, we provide mitigation strategies to thwart adversaries and directions for future DER cybersecurity research.}
\end{abstract}


%%%%%%%%%%%%%%%%%%%%%%%%%%%%%%%%%%%%%%%%%%%%%%%%%%%%%%%%%%%%%%%%%%%%%%%%%%%%%%%%
%%%%%%%%%%%%%%%%%%%%%%%%%%%%%%%%%%%%%%%%%%%%%%%%%%%%%%%%%%%%%%%%%%%%%%%%%%%%%%%%
%%%%%%%%%%%%%%%%%%%%%%%%%%%%%%%%%%%%%%%%%%%%%%%%%%%%%%%%%%%%%%%%%%%%%%%%%%%%%%%%
%%%%%%%%%%%%%%%%%%%%%%%%%%%%%%%%%%%%%%%%%%%%%%%%%%%%%%%%%%%%%%%%%%%%%%%%%%%%%%%%
%\input{1-Introduction}

\section{Introduction} \label{s:intro}

In the last decades, electric power systems (EPS) have undergone significant transformations (e.g., decentralized generation,  digitization of customer services, smart grid initiatives, etc.) to meet the increasing power demand and provide economically and environmentally friendly energy. Renewable energy resources (RES) harnessing wind, solar, and thermal energy have been used to improve energy efficiency while meeting stringent carbon emission regulations \cite{grid_mod}. The shift towards more sustainable grid architectures has also boosted the adoption of distributed energy resources (DER).
%%%% DER definition and numbers
According to the definition of the North American Electric Reliability Corporation (NERC), DERs are distribution-level resources that produce electricity and are not part of the bulk EPS \cite{NERCDER}. 

DERs can leverage renewable or non-renewable resources. % for power generation. Typical DER examples include, \emph{i)} rooftop solar photovoltaic (PV), \emph{ii)} wind (micro)turbines, \emph{iii)} diesel generators, \emph{iv)} microgrids, \emph{v)} controllable loads and demand-response, \emph{vi)} electric vehicles (EV) and electric transportation, and \emph{vii)} behind- or in front-of-the-meter energy storage (e.g., battery energy storage systems -- BESS, flywheels, etc.) \cite{NERCDER}. %connected on the distribution transformer secondary \cite{NERCDER}. 
DER devices are classified into different categories based on their operation principles, e.g., generation, energy storage, combination of the two, or controllable loads. Solar panels and wind turbines belong to the generation category \cite{sasan2017}, batteries and electric vehicles (EVs) fall into energy storage, combined cooling and heating, and electric water heaters are examples of controllable loads \cite{soltan2018blackiot}, and inverter-based resources (IBR) can be used in both generation and storage scenarios. \looseness=-1

The rapid integration of DERs highlights their importance for EPS and %The high demand for DERs has also resulted in a decrease in the production cost of solar cells and lithium batteries. %, which has expedited their adoption. 
%Studies predict that 
the global generation capacity is expected to grow from 132.4 GW in 2017 to 528.4 GW by 2026 \cite{DERgoals}. Similar observations can be made for battery energy storage systems (BESS). Namely, in the United States (U.S.), BESS are expected to grow from 1.2 GW in 2020 to nearly 7.5 GW in 2025, while developing into a market of \$7.3 billion annually \cite{utilitydive}. 
Strategic placement of DERs, can utilize on-site generation and minimize utility costs by deferring investments for the expansion of the power system network. DER-generated power does not need to be transferred from remote bulk generation facilities, minimizing energy losses and providing economically dispatchable power. DERs could reduce transmission system operator (TSO) ancillary market costs, which increased by almost 70\% during the 2020 COVID-19 lockdown \cite{ghiani2020impact}. This was due to the stochasticity of real-time power demand and the requirement to maintain frequency stability and energy reserves, mainly relying on the intermittent generation of RES, which created a very competitive power market. DERs can enhance EPS reliability through DER-supported autonomous functions. % -- if faults, blackouts, black swan, or other disruptive incidents occur. 
With higher DER penetration, the consequences of disastrous events %, e.g., extreme weather, cyberattacks, etc., 
could be averted or effectively mitigated \cite{syrmakesis2022classifying}. For instance, DERs could alleviate the impacts of extreme weather events similar to the Texas snowstorm in February 2021 \cite{Texas}.\looseness=-1
%(which caused 4.5 million homes to lose electricity, resulting in more than 57 deaths and \$195 billion in property damage) 

% DER potential security implications
The flexible and autonomous attributes of DERs render them invaluable pillars for the EPS critical infrastructure (CI). 
%With the term CI, we define any operational sector (e.g., public health, water, energy, transportation, telecommunications, etc.) which, if compromised, can jeopardize national security~\cite{lewis2019critical}.
%According to~\cite{lewis2019critical}, we can identify $11$ sectors of critical infrastructure and $5$ key assets. These CI sectors include, among others, public health, water, energy, transportation, telecommunications, etc., while the critical assets entail nuclear power plants, dams, government facilities, commercial key assets, and national monuments. 
%The importance of each sector -- and the priority in safeguarding it -- might have changed with the passage of time, however, energy systems and power system assets have always remained vital parts of the CI backhaul. 
Being crucial parts of the EPS, DERs can become prominent cyberattack targets. Attackers could exploit vulnerabilities, i.e., weaknesses in a system that can be leveraged to carry out an attack, to compromise remote communication, gain control of DERs, and propagate attack impacts to the rest of the system. 
%The effects of non-functional and/or maliciously controlled EPS can be dire. 
The cybersecurity incident in Utah, U.S. on 5 March 2019 is a prime example of how compromised communications can affect grid services. In this incident, attackers exploited vulnerabilities in security firewall devices to halt communications between system operators and distribution wind and solar utilities \cite{CVE-2018-0296}. The impact of this denial-of-service (DoS) attack caused loss of visibility of power grid assets and could have caused interruptions of electrical system operations \cite{learned2019risks}. %\cite{electricReport, learned2019risks}.
% which leads many security analysts in examining ways to efficiently detect, protect and mitigate such cyberthreats. 
To overcome cyberattacks comprehensive security investigations should be performed taking into account the \emph{cyber-physical} DER nature. \looseness=-1
% which could be targeted on either the \emph{cyber} or \emph{physical} threat plane.\looseness=-1

\begin{figure}[t]
    \centering
    \includegraphics[width=0.75\linewidth]{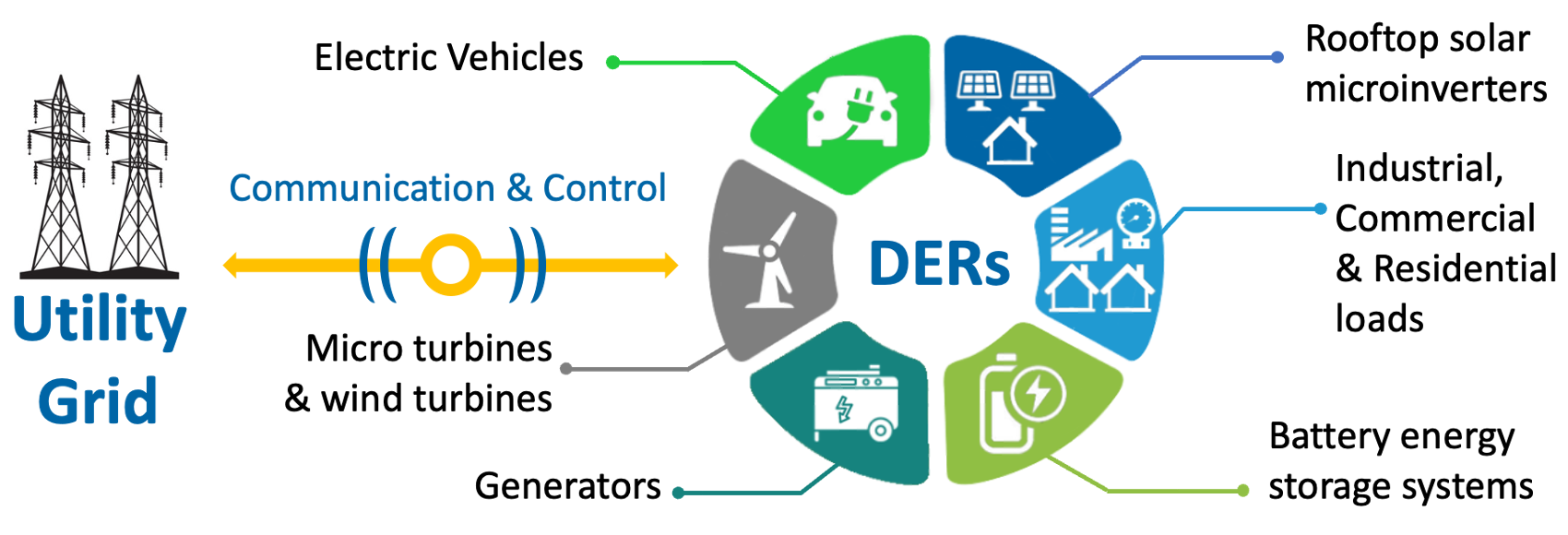}
    \vspace{-2mm}
    \caption{Utility-to-DER interconnection.}
    \label{fig:DERs}
    \vspace{-2mm}
\end{figure}

%DER protocol issues
On the cyber level, protocols facilitate the %for instance, to utilize the previously discussed DER advantages,
communication between DERs and utility aggregators for DER monitoring and control purposes (Fig. \ref{fig:DERs}). Information and communication technologies (ICT) enable communication between DER assets and DER management systems (DERMS). Typically, wired or wireless communication protocols are used, such as the IEEE 1815-Distributed Network Protocol (DNP3), Sunspec Modbus, open automated demand response (OpenADR), and IEEE 2030.5 (Table \ref{table:DER_coms}). Embedded devices on the DER side handle monitoring and remote DER asset control requests. Insecure remote communication expands the EPS threat surface since adversaries can exploit communication protocol weaknesses to mount attacks, e.g., DoS,  man-in-the-middle (MitM), etc. The situation is aggravated by the fact that many of the communication protocols have known vulnerabilities, which if exploited, % Exploiting those vulnerabilities 
can compromise system operation by issuing malicious commands to DER end-devices. % (Section \ref{sec:protocol_level}).

\begin{table}[t!]
\small
    \setlength{\tabcolsep}{1.2pt}
    \centering
    \caption{Common DER communication protocols. }
    \label{table:DER_coms}
    
    \renewcommand{\tabularxcolumn}[1]{m{#1}}

    \begin{tabularx}{\linewidth} { 
      || >{\hsize=.7\hsize\textwidth=\hsize\raggedright\arraybackslash}X 
      | >{\hsize=2.0\hsize\textwidth=\hsize\centering\arraybackslash}X
      | >{\hsize=.3\hsize\textwidth=\hsize\centering\arraybackslash}X || }
      
     \hline \hline
     {Protocol} & {{Description}} & {Std.} \\ 
     \hline
     {IEEE 1815 DNP3} & {Interoperable communication framework for secure information exchange in industrial systems (e.g., SCADA)}  & {\cite{1547std}}  \\ \hline
     
     {Sunspec Modbus} & {Modbus protocol extension for DER parameters (e.g., power, voltage) and ancillary services monitoring and control}  & {\cite{ModBus}}  \\ \hline
     
     {OpenADR} & {Energy market management standard regulating demand-response via signals to DERs and other controllable devices }  & {\cite{OpenADR_standard}}  \\ \hline
     
     {IEEE 2030.5} & {Smart energy profile application standard and default protocol for DER management   }  & {\cite{2030} }  \\ \hline
    
    \hline \hline
    \end{tabularx}

\end{table}

%DER device level issues
On the physical layer, the device-level consists of the embedded architectures (e.g., controllers, gateways, converters, etc.) and their fundamental components (e.g., hardware, firmware, software, etc.) that support DER operations and could constitute another weak link for the system security. % of the whole system. 
Most of these embedded devices are built using commercial off-the-shelf (COTS) components that could suffer from hardware- and software- level vulnerabilities. %As a result, embedded device-related vulnerabilities at the hardware and software level can be ported to DER systems. 
Apart from the vulnerabilities of COTS and the computational resource constraints of embedded systems (which limits the sophistication of security schemes), their trustworthiness cannot be attested either. The heterogeneity of embedded systems aggravates their security posture and makes verifying the supply-chain trustworthiness often infeasible. \textcolor{black}{Multiple third-party vendors provide intellectual property (IP) blocks which are then integrated by fabrication facilities during manufacturing \cite{intel}. Apart from hardware IP blocks, software supply-chain compromises, where threat actors intentionally plant backdoors into software products to weaponize them later, are also a major concern \cite{CISA_supply}. The severe impact of supply-chain attacks was demonstrated during the Solarwinds incident in December 2020, which targeted network management systems around the globe. Namely, {18$k$} of the {300$k$} Solarwinds customers were running vulnerable versions of the Orion platform, including the U.S. Department of Treasury, U.S. Department of Energy, U.S. Department of Defense, 425 of the U.S. Fortune 500, as well as the cybersecurity firm FireEYE \cite{solarwinds}. The severe blow of the Solarwinds supply-chain attack on CI networks forced the White House to issue an executive order instructing the improvement of cybersecurity in all federal government networks \cite{CSexecorder}.  %Although Solarwinds supply-chain compromise did not target physical devices, such incidents should not be limited to software. 
Supply-chain-related dependencies
%compromises in the device physical layer 
arise as potential threat vectors, i.e., specific paths or methods that can be exploited to compromise a system, especially given the lack of security controls when designing or integrating COTS software or hardware during system design. }

\begin{figure*}[t]
    \centering
    \includegraphics[width=0.95\textwidth]{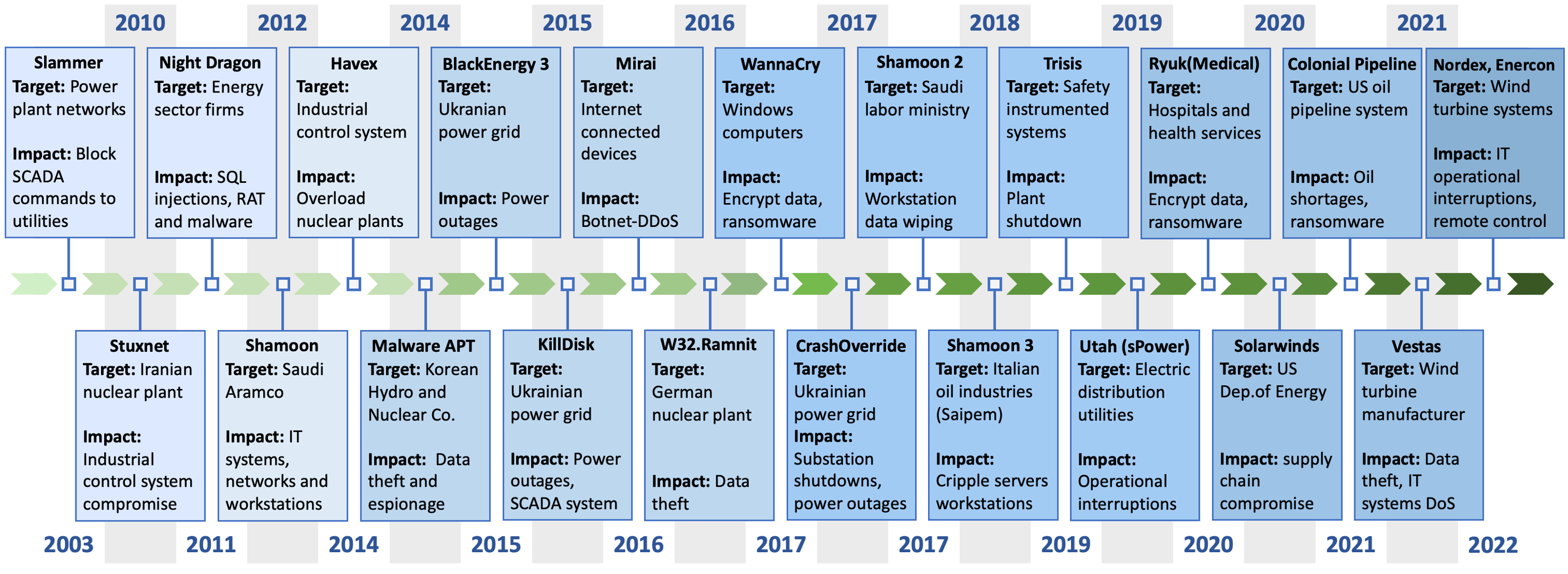}

    \caption{Timeline of cyberattacks targeting the energy sector and other critical infrastructure sectors.} 
    \label{fig:cyberattacks}
\end{figure*}

\begin{figure}[t]
    \centering
    \includegraphics[width=0.95\linewidth]{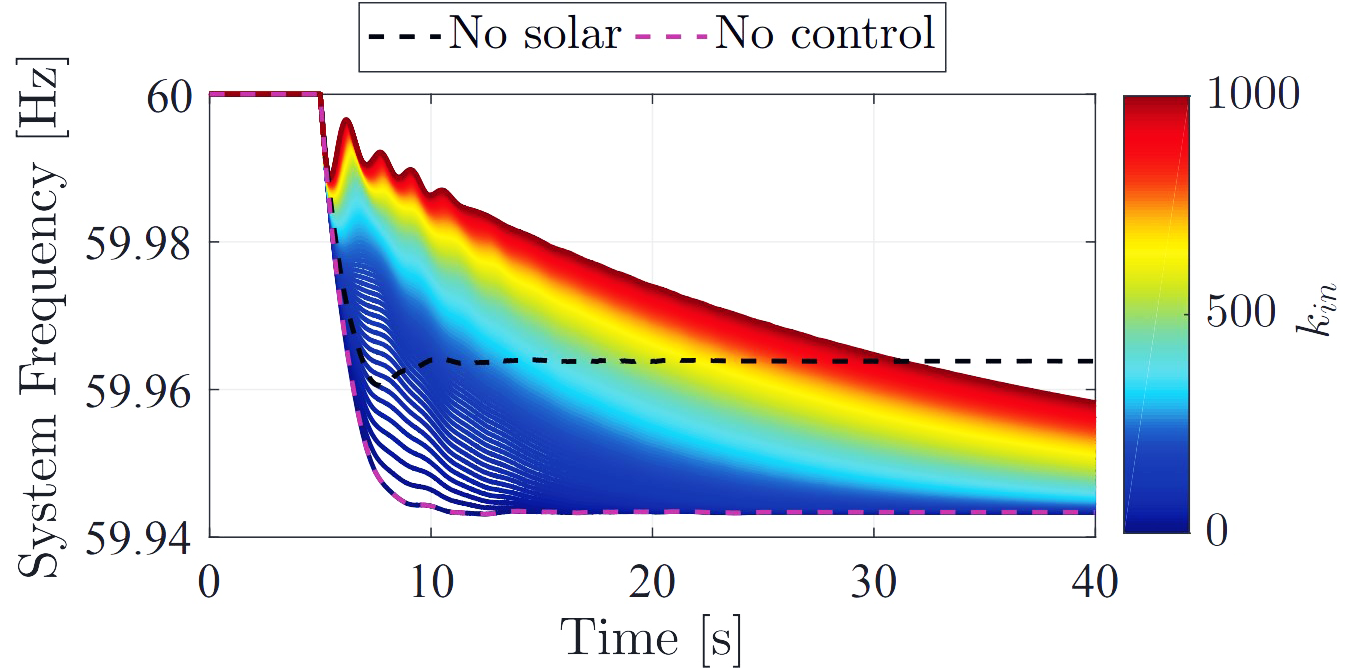}
    \caption{Impact of synthetic inertia gain ($k_{in}$) on system frequency response\cite{concepcion2017effects}.}
    \label{fig:inertia}
    \vspace{-5mm}
\end{figure}

Sophisticated cyberattacks targeting CI are gaining popularity given the severe impacts on business operations like the Solarwinds incident and the colonial pipeline ransomware attack, which halted the delivery of transport fuel to the Atlantic coast \cite{colonial}. Furthermore, attacks on public health are rising, such as the attack on the Florida water treatment plant \cite{florida}, attacks on hospitals during COVID-19 \cite{10.1093/intqhc/mzaa117}, etc.
%Worryingly, though, attacks targeting public health are also increasing, such as the attack on Florida's water treatment facility \cite{florida}, the attacks targeting hospitals during COVID-19 \cite{hospitals, 10.1093/intqhc/mzaa117}, etc. 
The aforementioned incidents underline that CI protection needs to become the epicenter of cybersecurity research. In Fig. \ref{fig:cyberattacks}, we demonstrate a timeline of cyberattacks targeting the energy and other CI sectors that have been reported in the literature. 
\textcolor{black}{The rapid penetration of inverter-based DERs has displaced synchronous generation arising concerns about the stability of IBR-dominated systems. Ancillary services, such as synthetic inertia, are crucial to maintaining frequency stability in grid and microgrid architectures composed of non-synchronous DERs \cite{concepcion2017effects}. However, from an adversarial perspective, these services can be exploited to destabilize such inertia-less systems, as demonstrated in Fig. \ref{fig:inertia}.} 
%As a result, RES and DERs are evidently becoming prominent targets. 

%This was also the case in \cite{learned2019risks}, where a firmware vulnerability in operational technology equipment (OT), specifically CISCO firewalls, was exploited to cause a DoS that disrupted power system communications and impeded the control of small renewable generation facilities. 

To combat threats related to DER, existing efforts such as the DER cybersecurity framework (DERCF) by the National Renewable Energy Lab (NREL) \cite{DERCF}, and the renewable energy and distributed systems integration (RDSI) program by Sandia National Lab (SNL) \cite{RDSI}, place grid security as one of their core initiatives. 
\textcolor{black}{Cybersecurity endeavors led by academic institutions have also attempted to identify vulnerabilities in DERs, and more specifically inverter-based systems. For instance, \cite{qi2016cybersecurity} and \cite{li2022cybersecurity} provide a detailed description of the smart inverter operational objectives and ancillary grid support functions. In \cite{li2022cybersecurity}, the authors discuss a variety of cyberattack and potential mitigation strategies. However, the connection between adversarial incentives and the resources necessary to perform such compromises is overlooked. In \cite{tuyen2022comprehensive, ye2021review} a comprehensive review of detection and defense methodologies is presented to overcome contingencies in photovoltaic (PV) systems. However, some of the prescribed defense strategies, e.g., blockchain, neglect the operational technology (OT) constraints and limitations of the underlying field devices.}

\textcolor{black}{Research efforts have also identified existing vulnerabilities in communication protocols, used for the orchestration of DER, that can be exploited as threat vectors \cite{DNP3_taxonomy, Huitsing2008AttackTF}. The standards defining the security objectives of such protocols are compiled in \cite{HASAN2023103540} along with cybersecurity principles that could help overcome communication-related threats. However, the performance overhead, associated with improving the security of such protocols, is not discussed. In \cite{johnson2020assessing}, the authors investigate the impact of securing the communication infrastructure on the real-time operation of DERs in a network co-simulation environment. Suggestions to overcome protocol cybersecurity limitations and metrics to assess cyberattack impacts are also reported, while simulation tools to measure the efficacy of protocol defenses are delineated in \cite{siqueira2020recommended}. }

\textcolor{black}{In this paper, we focus on cyberattacks targeting DER assets on both the \emph{device} and \emph{communication} levels. We discuss cyberattacks targeting the DER devices themselves and their % compromises  DERs play a significant role in reliable and resilient EPS operation, given their 
autonomous capabilities (e.g., defensive islanding) and ancillary services (e.g., voltage/frequency regulation, active/reactive power compensation, etc.). %\cite{ORNL}. 
DER security should be viewed holistically and is contingent upon the security posture of the inherent DER device architectures (e.g., vendor-specific), the utilized communication protocols, and control schemes (e.g., user, aggregator, or utility -defined). Comprehensive security solutions should encompass an \emph{adversary viewpoint} capable of not only mitigating previous incidents (as most of the literature does), but proactively defending against ``what could happen" scenarios. To bridge this overlooked research gap, we first discuss the motives and resources of adversaries and the crucial components of DER systems before we focus on DER attacks at the \emph{i)} communication protocol and \emph{ii)} device levels. Last, DER protocol- and device- level vulnerabilities, attacks, impacts, and mitigation strategies are presented. }

%\hl{This needs to be changed to reflect the changes in the structure of the paper and the additions of sections}

The roadmap of this work is as follows. Section \ref{s:threat_model} presents the adversary and attack models. Sections \ref{sec:protocol_level} and \ref{sec:device_level} outline the protocol- and device-level vulnerabilities, the corresponding cyberattacks, and potential impacts and mitigations. %are also reported. 
Section \ref{s:conclusion} concludes the paper and provides a brief discussion of DER cybersecurity metrics and future challenges. % that need to be addressed. \looseness=-1

% paper: Effects of Communication Latency and Availability on Synthetic Inertia

%%%%%%%%%%%%%%%%%%%%%%%%%%%%%%%%%%%%%%%%%%%%%%%%%%%%%%%%%%%%%%%%%%%%%%%%%%%%%%%%
%%%%%%%%%%%%%%%%%%%%%%%%%%%%%%%%%%%%%%%%%%%%%%%%%%%%%%%%%%%%%%%%%%%%%%%%%%%%%%%%
%%%%%%%%%%%%%%%%%%%%%%%%%%%%%%%%%%%%%%%%%%%%%%%%%%%%%%%%%%%%%%%%%%%%%%%%%%%%%%%%
%%%%%%%%%%%%%%%%%%%%%%%%%%%%%%%%%%%%%%%%%%%%%%%%%%%%%%%%%%%%%%%%%%%%%%%%%%%%%%%%
%\input{2-ThreatModel}

\section{Threat Modeling}\label{s:threat_model}
This section identifies high-value targets in DER-integrated systems and attack vectors with their corresponding cyber-threats. 
\textcolor{black}{Assumptions about adversary capabilities and knowledge, and attack specifics following the threat model methodology are introduced in \cite{zografopoulos2021cyber}. According to MITRE's ATT\&CK for industrial control systems (ICS) framework the distinction between the \emph{adversary} and \emph{attack} models is crucial. The former allows us to evaluate a threat incident from the \emph{``what could happen''} attacker viewpoint, instead of the \emph{``what did happen''} defender's angle. The attack model, describes the requirements for a vulnerability to become a system threat \cite{MITRE_ICS}. We compile the threat vectors and the methodology followed to materialize attacks (Table \ref{tab:vectors}), and provide cyberattack definitions %, (Table \ref{tab:common_attacks}), 
which are used in Sections \ref{sec:protocol_level} and \ref{sec:device_level}.}

%%%%%%%%%%%%%%%%%%%%%%%%%%%%% BEGINNING OF TABLES %%%%%%%%%%%%%%%%%%%%%%%%%%%%%

\begin{table*}[t]
\small
\setlength{\tabcolsep}{1.2pt}
\centering

\caption{\textcolor{black}{Attack vector description and potential threats for DER assets.} }

\label{tab:vectors}

\renewcommand{\tabularxcolumn}[1]{m{#1}}

\begin{tabularx}{\textwidth} { 
  || >{\hsize=.4\hsize\linewidth=\hsize\raggedright\arraybackslash}X 
  | >{\hsize=1.8\hsize\linewidth=\hsize\centering\arraybackslash}X
  | >{\hsize=.8\hsize\linewidth=\hsize\centering\arraybackslash}X || }
  
 \hline \hline
 {Attack vector} & {{Description}} & {Threat} \\ 
 \hline
 
 {Lack of interoperability} & {DER architectural diversity and implementation specifications (e.g., security requirements) can result in inter-system insecure communications.}  & {DER denial of legitimate messages and control commands}  \\
\hline

{Data integrity violations} & {Stored, transmitted or received data is modified without validation, causing DER malfunction or allowing unauthorized access to control/log information.}  & {Malicious modification of control parameters}  \\
\hline

{Implementation errors} & {Security flaws within systems and/or communication modules enabling the remote control of DER assets and exfiltration of historical generation data.}  & {Command and control of load/demand-side devices}  \\
\hline

{Supply-chain compromises} & {Installment of malicious hardware-based eavesdropping programs, worms, oversights during manufacturing of components, devices, or systems. }  & {Sensitive information disclosure}  \\
\hline

{Insecure firmware} & {Digital signatures of firmware updates are not verified, granting malware (viruses, worms, trojans, etc.) access to otherwise secure systems.}  & {DER systems privilege escalation}  \\
\hline \hline
\end{tabularx}

\end{table*}

%%%%%%%%%%%%%%%%%%%%%%%%%%%%% END OF TABLES %%%%%%%%%%%%%%%%%%%%%%%%%%%%

\subsection{Adversary Model} \label{sec:attack_assump}
\textcolor{black}{The following discussion delineates the assumptions followed in this study regarding the adversarial system knowledge and capabilities, and how they correlate with attacks targeting DER grid communications and devices. Adversaries could rely on publicly available open-source intelligence information to perform their attacks \cite{acharya2020public}.} Additional knowledge can be acquired after a system asset or device is compromised and/or while a cyberattack is propagated in the system. %Attack trees are conceptual representations which illustrate the ways that an adversary can exploit a system asset and acquire system information or compromise it. 
Knowledge can also be obtained by accessing unsupervised hosts on enterprise networks \cite{kang2015investigating,siqueira2020recommended}, by gaining unauthorized access to data shared among DER devices. 
\textcolor{black}{Other methods include eavesdropping, intercepting exchanged data \cite{zhou2019secure}, IP identification and port scanning, application-level protocol exploitation \cite{kang2015investigating, teymouri2018cyber}, ciphertext decryption, or  malicious insiders. \looseness=-1}

Equally important to the attacker knowledge assumptions are the adversarial capabilities (i.e., \emph{``what could happen''}). Adversarial capabilities consider the access to DER assets within a given system. For example, an attacker might be able to connect to remote DER devices through legitimate Bluetooth or speedwire connections (i.e., insider case) \cite{sundararajan2018survey}. Additionally, attackers could possess or have physical access to EV charging stations or to the local area network (LAN) over which DERs communicate \cite{asal2018}. In the latter case, such access could be achieved if attackers can infiltrate the LAN using proxy attacks on PCs, routers, surveillance systems, etc. \cite{sebastian2017exploring}. Additionally, other attacks assume that the adversary is capable of compromising peripheral controllers and deploying custom firmware \cite{kuruvila2021hardware, zografopoulos2022time} (e.g., change voltage measurement values in human-machine interfaces (HMIs) \cite{teymouri2018cyber}). % which may not comply with the power grid security standards. 
As a result, attackers can remotely manipulate and jeopardize various physical assets (e.g.,  distribution level devices, DERs, substation equipment, etc.) \cite{acharya2020public}. Attackers are assumed to have sufficient computational resources -- especially when supported by criminal organizations or nation-state actors -- to crack passwords and decrypt power grid data. For example, it is reported that by brute forcing PIN codes of wind farm control panels, attackers could send malicious commands to wind turbines impacting grid operation \cite{asal2018}.

\subsection{Attack Model} \label{sec:attack_model}

The attack model specifics are essential for the threat model description (e.g., in retrospection of past cyberattack incidents). Although we do not provide one-to-one mappings between attacks (presented in this study) and each corresponding attack model component, their key aspects -- influencing impacts and mitigation strategies -- are described in Sections \ref{sec:protocol_level} and \ref{sec:device_level}. Attack model information is crucial when establishing the requirements for a vulnerability to develop into a system compromise, its potential impact, and mitigations or other methods to overcome such adversity. Following the attack methodology in \cite{zografopoulos2021cyber}, the attack model can be considered as being composed of the following six elements, \emph{i)} the attack frequency, \emph{ii)} the attack reproducibility, \emph{iii)} the functional level being attacked in the system, \emph{iv)} the attacked asset, \emph{v)} the techniques being used, and \emph{vi)} the attack premise. For instance, some attacks might have to be performed iteratively and reproduced multiple times to compromise the system behavior. This could be the case with packet replay, DoS, and MitM attacks, where attacks might aim to spoof DER communications or exhaust DER protocols and/or device resources causing intermittent, slow, or loss of communication thereof \cite{carter2017cyber}. The attack functional level and the targeted DER assets notably differ when considering communication protocol and device attacks. Typically, cyberattacks targeting DER devices could be assigned to levels 0 and 1 including process sensors, actuators, and controllers, while attacks on the communication, coordination, and control fabric of DER systems target the higher levels (i.e., 3, 4, and 5) of the Purdue model \cite{ackerman2017industrial}.

Distinctions are also necessary when investigating the attack techniques and the premise of cyberattacks. A detailed presentation of the most common and recent attack techniques used by adversaries can be found in \cite{MITRE_ICS}. The attack premise delineates whether the attack is performed in the cyber or physical system domain. For our study, identifying the attack premise is fairly intuitive since DER protocol attacks are limited to the cyber domain, while DER device compromises could span both the cyber and the physical domains. Information on how complex cyberattacks could target the cyber and physical domains while exploiting diverse attack techniques is provided in \cite{zografopoulos2021cyber}. Specifically, in the case studies presented in \cite{zografopoulos2021cyber}, physical devices, e.g., inverter controllers, are compromised by exploiting control logic modification techniques impacting power conversion, power factor, active/reactive injections, etc. Additionally, cyberattacks targeting communications -- via time-delay attacks on the cyber domain -- and the impact of intercepted, delayed, and/or modified control commands on the simulated power system models are also demonstrated.

\subsection{DER Targets and Cyber-Threats} \label{sec:attack_targets}
We refer to mission-critical system assets which can jeopardize grid operation if maliciously compromised by threat actors as \emph{crown jewels} \cite{MITRE_CJA}. Crown jewels span the ICT infrastructures, such as the stakeholder-to-DER device communication channels \cite{carter2017cyber}, physical-interfaces \cite{sebastian2017exploring}, and the DER devices themselves. Notably, DER devices include PV inverters or smart inverters \cite{teymouri2018cyber}, BESS, EVs, wind turbines \cite{asal2018}, demand-side loads \cite{cardenas2020assessing}, and DER controllers. Gaining access to any crown jewel, enables adversaries to manipulate DER power output (e.g., generated, stored, etc.), which can result in possible brownouts, false trips, feeder overloadings, voltage/frequency violations, damaged protection equipment or system instabilities \cite{qi2016cybersecurity}. %In Table \ref{tab:vectors}, we present a summary of attack vectors targeting DERs and inherent cyber-threats based on related security research studies  \cite{lai2017cyber, qi2016cybersecurity, soltan2018blackiot}.

We distinguish between attacks targeting the DER communication protocols used for the control and coordination of DER assets, and the actual DER embedded devices and their components, i.e., hardware, and software. Although the threat surface for these two categories might share similarities, it also has differences (e.g., attack access, exploitation tactics and techniques, etc.). Along with these differences, the interoperable nature of DER systems mandates comprehensive security mechanisms that treat the DER crown jewels jointly. %In Section \ref{sec:protocol_level} and \ref{sec:device_level}, we present the respective vulnerabilities, attacks, impacts, and potential mitigation and best practises when designing DER systems. 
Table \ref{tab:common_attacks} compiles common cyberattack definitions relevant to DER asset vulnerabilities. % (either on the protocol or device level).

\begin{table*}[t]
\small
\centering
\caption{ \textcolor{black}{Definitions of cyberattacks tailored for DER assets.} }

\label{tab:common_attacks}

\renewcommand{\tabularxcolumn}[1]{m{#1}}

\begin{tabularx}{\textwidth} { 
  || >{\hsize=.4\hsize\linewidth=\hsize\raggedright\arraybackslash}X 
  | >{\hsize=2.25\hsize\linewidth=\hsize\centering\arraybackslash}X
  | >{\hsize=.35\hsize\linewidth=\hsize\centering\arraybackslash}X || }
 \hline \hline
 {Attack Type} & {{Definition}} & {Reference}  \\ 
 \hline
 
 {Network Reconnaissance} & { Adversary performs vulnerability scanning on a network. Information such as IP/MAC addresses of DER devices, open ports, services running, and type/version of the system can be disclosed.} & {\cite{carter2017cyber}} \\
\hline
{Eavesdropping} & { Attacker “listens” to confidential in-transit data potentially stealing sensitive information.} & {\cite{henry2015cyber}, \cite{onunkwo2018cybersecurity}}  \\
\hline
{Man-in-the-Middle (MitM)} & { Adversary redirects communication through compromised ``middle'' node (e.g., network switch, gateway, etc.) enabling packet monitoring and modification before they reach their destinations.} & {\cite{siqueira2020recommended, onunkwo2018cybersecurity, soltan2018blackiot}}  \\
\hline
{Denial of Service (DoS)} & { Target resources are overloaded (ports are flooded with traffic), inhibiting its nominal operation. Targeted devices become ``laggy'', unable to timely issue responses to legitimate device requests.} & {\cite{carter2017cyber}}  \\
\hline
{Packet Replay} & { Data transmissions between DER client applications and the DER devices are recorded, modified, and retransmitted by attackers at different time instances.} & {\cite{carter2017cyber}, \cite{lai2017cyber}}  \\
\hline
{Supply-chain} & { Adversary adds malware during the manufacturing, system integration, shipping, or installation stages. The malware can be weaponized remotely to perform unauthorized/unintended actions.} & {\cite{siqueira2020recommended}}  \\
\hline
{Brute forcing and fuzzying} & { Adversary attempts to guess credentials, encryption keys, etc. or bring the system to an unexpected state. Once successful, he/she could attain unauthorized access to system resources.} & {\cite{lai2017cyber}}  \\
%\hline
% {Wireless communications attack} & {\footnotesize An attacker taking advantage of devices that have wireless capabilities and is able to add, modify, and/or delete data wirelessly from a remote location bypassing any physical security protocols that have been put in place} & {\cite{soltan2018blackiot}}  \\

\hline \hline
\end{tabularx}
\end{table*}

%%%%%%%%%%%%%%%%%%%%%%%%%%%%%%%%%%%%%%%%%%%%%%%%%%%%%%%%%%%%%%%%%%%%%%%%%%%%%%%%
%%%%%%%%%%%%%%%%%%%%%%%%%%%%%%%%%%%%%%%%%%%%%%%%%%%%%%%%%%%%%%%%%%%%%%%%%%%%%%%%
%%%%%%%%%%%%%%%%%%%%%%%%%%%%%%%%%%%%%%%%%%%%%%%%%%%%%%%%%%%%%%%%%%%%%%%%%%%%%%%%
%%%%%%%%%%%%%%%%%%%%%%%%%%%%%%%%%%%%%%%%%%%%%%%%%%%%%%%%%%%%%%%%%%%%%%%%%%%%%%%%
%\input{3-Protocols}

\section{DER Protocol Level} \label{sec:protocol_level} %\hl{inro sentence here missing}
\textcolor{black}{The following section elucidates DER protocol-level security oversights and furnishes approaches to mitigate and overcome the impact of intrusions exploiting the communication plane. We systematically examine prominent industrial control protocols and demonstrate the DER cyber kill chain, indicating the steps adversaries follow to compromise EPS operations. \looseness=-1}

\begin{comment}

\begin{figure*}[t]
\centering
    \subfloat[]{
        \includegraphics[width=0.18\textwidth]{figures/interrupt.png}
        \label{fig:interrupt}
    } \hspace{4mm}
    \subfloat[]{
        \includegraphics[width=0.18\textwidth]{figures/intercept.png}
        \label{fig:intercept}
    } \hspace{4mm}
    \subfloat[]{
        \includegraphics[width=0.18\linewidth]{figures/modify.png}
        \label{fig:modify}
    } \hspace{4mm}
    \subfloat[]{
        \includegraphics[width=0.18\linewidth]{figures/fabricate.png}
        \label{fig:fabricate}
    }\\
\vspace{-2mm}  
\caption[CR]{DER protocol level attack categories including data and/or command \subref{fig:interrupt} interruption (e.g., sinkhole), \subref{fig:intercept} interception (e.g., port stealing), \subref{fig:modify} modification (e.g., ARP poisoning), and \subref{fig:fabricate} fabrication (e.g., MAC- and SYN-flooding) attacks.} 
\vspace{-3mm}
\label{fig:attack_types}
\end{figure*}

\end{comment}

\subsection{DER Protocol Level Vulnerabilities} \label{sec:protocol_vuln}
\begin{figure}[t]
\centering
\includegraphics[width=0.7\linewidth]{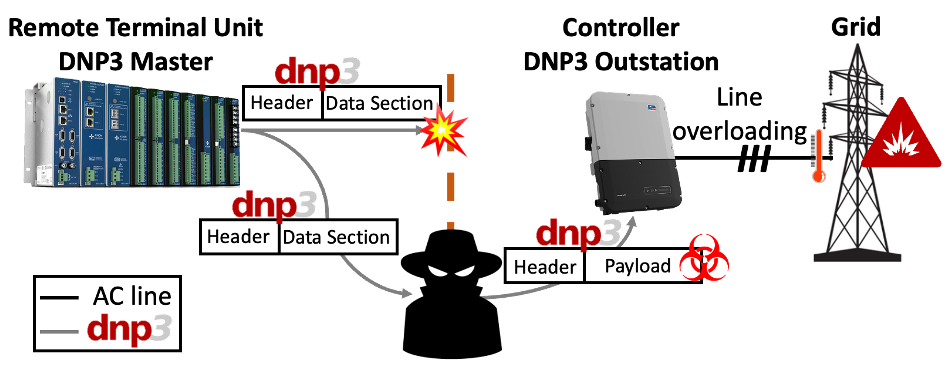}
\caption{DNP3 attack on DER controller: attacker intercepts the control command (from master) and issues a maliciously fabricated command to the outstation device.}
\label{fig:DNP3_attack}
\vspace{-2mm}
\end{figure}

\textcolor{black}{The most commonly used protocols for DER communication include the IEEE 1815-DNP3, Modbus, OpenADR, and IEEE 2030.5 \cite{Johnson2017Roadmap}. However, most of these protocols did not originally support any overarching cybersecurity requirements \cite{PROTOS}. Specifically, Modbus and DNP3, two widely used serial protocols in process automation systems,  have several identified vulnerabilities. %\cite{drias2015taxonomy}. 
For instance, the authors in \cite{DNP3_taxonomy}, provide a comprehensive attack taxonomy for DNP3 and discuss 28 different attack types. 
The presented attacks can be classified into four broad threat categories, including DNP3 data \emph{i)} interruption, \emph{ii)} interception, \emph{iii)} modification, and \emph{iv)} fabrication attacks. % (Fig. \ref{fig:attack_types}). 
Interruption and interception attacks could delay control commands to industrial assets or acquire data in-transit. However, modification and fabrication attacks arise as the most threatening since they can send erroneous data to affect industrial processes, alter device configurations, and spoof master controllers, as shown in Fig. \ref{fig:DNP3_attack}. The attack impacts range from the acquisition of network and device configuration data to the corruption of remote devices and acquiring control of DNP3 master units. Similarly, more than 700 vulnerabilities have been identified in the open source CERT Java-based implementation of OpenADR \cite{park2014research}. Some of those vulnerabilities could be remotely exploited, compromising sensitive data such as usernames, system properties, installation directories, etc \cite{CVE-2010-3860}. }

\textcolor{black}{To enhance the security of DER communication, IEEE 1547-2020 interconnection and interoperability standard identifies the security requirements to be satisfied by both communicating parties \cite{IEEE1547}. Although compliance with IEEE 1547 can safeguard the exchanged data and control commands issued between DERs and aggregators, cyberattacks aiming to disrupt grid operations (e.g., DoS) remain. Of the aforementioned DER protocols, only IEEE 2030.5 was developed with stringent cryptography requirements \cite{2018ieee}. The latest versions of DNP3 \cite{IEEE1815}, Sunspec Modbus 700 series \cite{SunspecModbus}, and OpenADR 2.0 conform with IEEE 1547-2020 and National Institute of Standards and Technology (NIST) requirements. As such, they can be seen as semantically identical from a security standpoint with IEEE 2030.5 \cite{SunspecModbus}. Furthermore, transport layer security (TLS 1.2 or 1.3) is advised for all protocols to enhance wireless communications security between aggregators and DER edge devices \cite{lai2017cyber, DNP3-TLS}.}

\textcolor{black}{Although there are secure communication protocols, the network communication infrastructure remains a prominent threat vector for adversaries. % can exploit to mount their attacks. 
This can be attributed to the hesitation of industrial facilities to switch to newer and more secure communication protocols considering the potential economic or operational overheads. The plethora of legacy devices is also a limiting factor that restricts the modernization efforts of the communication fabric. The situation is further exacerbated since according to the Electric Power Research Institute (EPRI), more than 75\% of North American electric utilities use the DNP3 protocol for industrial control applications and supervisory control and data acquisition (SCADA) systems \cite{DNP3_attack}. For instance, attacks targeting DNP3 communications could either aim to exploit vendor implementations of the protocol, protocol specifications, or vulnerabilities in the supporting communication infrastructure \cite{DNP3_taxonomy}. In addition to DNP3 protocol vulnerabilities, user datagram protocol (UDP) vulnerabilities arise since, in some systems, portions of DER modification requests are passed in plaintext \cite{qi2016cybersecurity}, which presents a critical control vulnerability to the DER system. }%Not to mention that multiple EPS assets might be still using outdated/vulnerable version of the aforementioned protocols since updating every node and endpoint of the communication fibre might be infeasible.

\textcolor{black}{Applying the latest  protocol versions using TLS and cryptographically secure communication channels can potentially deter adversaries. However, the security of the power grid system is contingent upon each weakest link. Knowing that many already deployed DER devices support insecure and outdated versions of the aforementioned communication protocols, limited computational capabilities prohibiting the implementation of sophisticated encryption algorithms or even updating their firmware (e.g., could be infeasible due to field operational constraints), urges for potent attack detection and mitigation mechanisms. Following, we discuss some of the eminent cyberattacks exploiting the pitfalls of protocol vulnerabilities.}

\subsection{DER Protocol Level Attacks} \label{sec:prot_attack}

To identify which layer of the protocol stack is targeted by each attack, the Open Systems Interconnection (OSI) basic reference model is used  \cite{zimmermann1980osi}. The OSI model has seven major layers: application, presentation, session, transport, network, data link, and physical layer. In this paper, we focus on the security of commonly used industrial control protocols and specifically consider the data link, network, and transport layers of the OSI stack. The physical layers as well as the session, presentation, and application layers are not discussed since no security requirements are defined for the physical layer (bit transmission layer), while the aforementioned higher OSI layers can be application-specific or supported by other application protocols (e.g., secure file transfer protocol -- FTP/SFTP, simple network management protocol -- SNMP, telnet, etc.). \looseness=-1

\textcolor{black}{Data link layer uses the Ethernet protocol which is vulnerable to media access control (MAC) spoofing attacks. MAC addresses can be spoofed, allowing Ethernet frames to be forwarded to adversaries \cite{sundararajan2018survey}. In addition, MAC flooding attacks target the MAC address tables used by switches to store the information of legitimate devices, and the specific ports to which each device is connected \cite{sundararajan2018survey}. 
On the network layer, the cybersecurity of DER device communication is specified by IEEE Std. 2030.5, which operates over the UDP and transmission control protocol (TCP) with support for IPv4 and IPv6 protocols \cite{lai2017cyber}. Many owners of DER devices can remotely communicate with their DER devices and receive data such as measurement statistics, network communication analysis, firmware updates, and more. }

Especially for Modbus and DNP3, multiple attacks have been reported that compromise the confidentiality, integrity, or availability of in-transit data. In \cite{Huitsing2008AttackTF} 28 attacks targeting Modbus TCP packets and 20 attacks targeting Modbus serial instances are reported. Threat actors with network access, can intercept, interrupt, modify and fabricate Modbus control packets causing DoS, and/or injecting bad data and malicious commands resulting in loss of situational awareness and asset controllability. Similar attacks on DNP3, i.e., packet interception, fabrication, etc., are reported in \cite{DNP3_taxonomy} focusing mainly on the data link and transport layers, achieving loss of confidentiality, awareness, and control of industrial assets. In \cite{kang2015investigating}, the authors demonstrate the feasibility of attacks that corrupting manufacturing message specification (MMS) communications (facilitated by IEC 61850) by reporting false inverter limits curtailing the power generation of DER assets.

%The client-to-DER device communication utilizes the UDP protocol. 
Attacks targeting UDP vulnerabilities, during client-to-DER device communications, such as packet replay attacks have also been reported \cite{qi2016cybersecurity}. During a successful packet replay attack, the plaintext requests are captured while being transmitted. % over a channel. 
The attacker can then send the captured packets and issue malicious commands to the DER, resulting in device malfunctioning. For IPv4/IPv6 protocols, possible attacks include, network reconnaissance, packet replay, MitM, and DoS \cite{carter2017cyber}. Sophisticated MitM attacks on client applications and DER device communications have multiple steps. First, the adversary eavesdrops on the DER-to-client communications and then performs address resolution protocol (ARP) poisoning and port stealing attacks \cite{carter2017cyber}. ARP poisoning forces the MAC address of the adversary to be linked to the IP address of the target. This technique enables the interception of data in-transit. Port stealing enables the interception and modification of data by the adversary before it is delivered to the destination. 

On the transport layer, the predominant attack is synchronization (SYN) flooding. SYN flooding is a type of DoS attack where the adversary sends continuous SYN requests to multiple ports of the target device using fake IP addresses \cite{sundararajan2018survey}. The target device sends acknowledgment packets (SYN-ACK) to each fake IP request.  Since no consequent actions are performed by the fake IP client, the target device's port remains open until the connection times out. % and another fake SYN request is received. 

%a type of DoS attack limiting the availability of the target device. 

\subsection{DER Protocol Level Impacts} \label{sec:prot_impact}

The MAC spoofing attack on the data link layer aims to give adversaries unauthorized access to the network by maliciously modifying the source IP address \cite{sundararajan2018survey}. Another attack on this layer, MAC flooding, can target DER devices such as inverters, which can result in  memory overuse and potential communication bottlenecks. The adversarial impact could lead to loss of availability in communication with the DER device, including the inability to control DER power management parameters. \looseness=-1

MitM and packet replay attacks on the network layer of DER device communication leverage the commands received from client applications using UDP/IP transport protocols. Such interception has been reported in \cite{carter2017cyber}, where sensitive information involving DER generation data and control commands were exchanged in plaintext format. Traffic analysis and in-transit packet inspection can be used to exfiltrate setpoint values. The underlying impact of a successful packet replay attack is demonstrated in \cite{sasan2017}, where the replay attack doubles the magnitude oscillations of the DER's real output power. In microgrids, especially during autonomous operation, such disturbances could result in relay trippings, damaged equipment, and potentially load shedding events \cite{8242351, peng2019stability, barua2020hall}. 

\textcolor{black}{The impacts of protocol-based DoS attacks, e.g., during SYN flooding, could diminish the availability of DER devices. Reference values for real/reactive power and phase measurements are critical for the operation of inverters. If such information is delayed, real-time reference values cannot be updated, and thus the DER operation is indicated by firmware-defined defaults \cite{sasan2017}. Such distortions can result in under- or over-generation, leading to uneconomic operation, and real or reactive power instabilities in the DER connected grid \cite{ospina2020trustworthy}.} 

\subsection{DER Protocol Level Mitigations} \label{s:prot_mitigation}
\textcolor{black}{Different mitigation strategies for protocol-level attacks have been reported in the literature \cite{sundararajan2018survey, zografopoulos2020special}. To mitigate spoofing attacks on the data link layer, authentication-based access control is necessary. In the presence of authentication-based access control, the communication between the client and the device has to be first authenticated before the client is allowed to connect and exchange data and control commands. An alternative proposed mitigation approach considers the network switch configuration, i.e., white-listing a limited number of trusted MAC addresses, while discarding requests from unknown sources. Network switch best practices suggest the deployment of an authentication, authorization, and accounting server, that manages connections and certifies that the MAC addresses are added to the table only after being authenticated.} 

On the network layer, the use of firewalls, one-way communication diodes, packet filters, circuit-level gateways, proxy servers, and two-way authentication can be utilized to prevent network reconnaissance, packet replay, MitM, and DoS attacks. The use of physical and logical network segmentation and perimeter security defenses can be used to prohibit access to critical parts of the system and prevent the lateral movement of threat actors between IT and ICS networks (if the prior is compromised) \cite{NISTSP800}. Demilitarization zones should also be enforced to aid network segregation and serve as proxies, avoiding the security hazards that network-connected devices could port to DER control functionalities. DER applications using TCP/IP are more resistant to packet replay attacks since unique session IDs are generated during the initial three-way TCP handshake (communication initialization) \cite{carter2017cyber}. TCP/IP transport includes additional IP header information that DER applications can use to prevent packet replay attacks. Leveraging the unique IP headers, the server is able to detect and drop duplicate packets, hence preventing packet replay attacks. \looseness=-1

Furthermore, the implementation of cryptographic techniques, secure key distribution, and exchange schemes can help prevent MitM and replay attacks \cite{zografopoulos2020derauth}. \textcolor{black}{Currently used cryptographic mechanisms might no longer be secure once quantum computers become widely available, and EPS stakeholders will have limited time to adapt their systems \cite{ahn2021quantum}. To address this issue, quantum key distribution (QKD) schemes have recently been proposed to improve the security of cryptographic keys \cite{kong2022review, tang2020quantum}. Contrary to other application fields, e.g.,  computer networks, online banking, etc., the power systems community has only recently started considering such approaches, and this can be mainly attributed to the fact that QKD schemes cannot be directly applied to deployed and legacy systems. In addition, testbeds to evaluate the real-time  performance of such quantum schemes do not exist. However, in the future, it is evident that quantum-secure encryption algorithms, QKD schemes, and the essential infrastructure to test them would be part of EPS cybersecurity research \cite{PQM}.\looseness=-1}

\begin{figure*}[t!]
\centering
\includegraphics[width=0.85\textwidth]{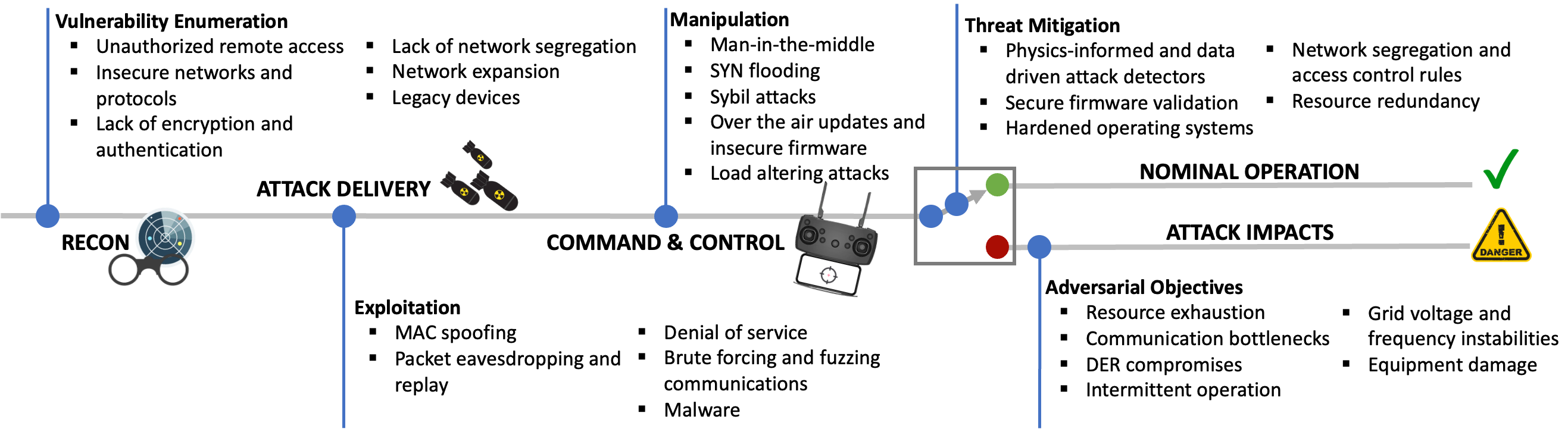}
\caption{DER cyber kill chain including communication layer vulnerabilities, attacks, impacts, and mitigations.}
\label{fig:killchain}
\vspace{-2mm}
\end{figure*}

On the transport layer, mitigation of SYN flooding attacks is achieved by cryptographic hashing and stack tweaking, which reduce the connection request timeout period dropping incomplete sessions. Cryptographic hashing will determine the legitimacy of the connection by sending the SYN-ACK packet with a code derived from the client's IP address, port number, and unique ID number. Dropping incomplete connections in stack tweaking frees up ports for legitimate connections. Firewalls, intrusion detection and prevention systems (IDS/IPS), and traceback and push-back services can limit excessive traffic, therefore countering DoS attackers. Fig. \ref{fig:killchain} presents the DER cyber kill chain, where we recapitulate protocol vulnerabilities and attack entry points, % discussed in Sections \ref{sec:protocol_vuln} - \ref{sec:prot_mitigation}, 
enumerate potential attack impacts, and mitigations to overcome adversities.

%%%%%%%%%%%%%%%%%%%%%%%%%%%%%%%%%%%%%%%%%%%%%%%%%%%%%%%%%%%%%%%%%%%%%%%%%%%%%%%%
%%%%%%%%%%%%%%%%%%%%%%%%%%%%%%%%%%%%%%%%%%%%%%%%%%%%%%%%%%%%%%%%%%%%%%%%%%%%%%%%
%%%%%%%%%%%%%%%%%%%%%%%%%%%%%%%%%%%%%%%%%%%%%%%%%%%%%%%%%%%%%%%%%%%%%%%%%%%%%%%%
%%%%%%%%%%%%%%%%%%%%%%%%%%%%%%%%%%%%%%%%%%%%%%%%%%%%%%%%%%%%%%%%%%%%%%%%%%%%%%%%

\section{DER Device Level} \label{sec:device_level}

\textcolor{black}{Different DER categories can be identified based on their operation principles, %, e.g., generation, energy storage, hybrid, controllable loads, etc. 
thus, DER device compromises can impact grid operations to varying degrees based on their category and system utilization. This section focuses on vulnerabilities, attacks, impacts, and mitigations on the DER device-level. }% considering the underlying physical components (e.g., controllers) and the inherent hardware, firmware, and software. \looseness=-1 }

%DER devices can be classified into different categories based on their operation principles, e.g., generation, energy storage, combination of the two, and controllable loads.
%For example, solar panels and wind turbines are examples of DERs belonging to the generation category \cite{sasan2017}, batteries and electric vehicles fall into energy storage, combined cooling, heating, and electric water heaters are examples of controllable loads \cite{soltan2018blackiot}, and smart inverters can be used in both generation and storage setups. 

\subsection{DER Device Level Vulnerabilities} \label{s:device_vuln}

Inverter-based resources are rapidly deployed in EPS on a commercial and residential scale. However, they remain vulnerable to cyberattacks that can modify or erase their default DER settings \cite{cardenas2020assessing}. Rooftop PV inverter control, for instance, is regulated both on the local level (primary) as well as at a higher level, i.e., secondary control. The primary control system is responsible for matching the inverter-generated energy to the consumer power demand. \textcolor{black}{Contrary to the local control system which operates on the consumer side, centralized secondary control resides on the concentrator/aggregator and utility sides. The secondary control system orchestrates inverter operation on a larger scale, by managing ancillary services and providing setpoint updates to local control systems regulating frequency and voltage deviations after a perturbation, e.g., load change, fault, etc., has occurred.} %optimally dispatching power and meeting dynamic power demands economically.
\textcolor{black}{Typically, these control systems are operated remotely, using Internet-of-Things (IoT)-based and third-party applications, which could expose such DER devices to cyberattacks. For example, in \cite{carter2017cyber}, the authors demonstrate that by eavesdropping DER communications (performed over the Internet), passwords, user data, serial IDs and device names can be extracted using network traffic analysis software (e.g., Wireshark). Furtheremore, in some cases, sensitive information is exchanged in plaintext format, highlighting the importance of end-to-end data encryption and endpoint security hardening on the physical device itself.}

\textcolor{black}{Similar security concerns exist for wind turbines which represent another significant DER type. %being rapidly integrated in the electric grid. 
Wind turbine control panels (WTCP) are used to control turbines and monitor their real-time operating conditions, e.g., their generation setpoints are dynamically controlled to meet the electricity demand.  
According to \cite{asal2018}, in some deployments, the only security practices employed against adversaries attempting to access the WTCPs are software-based credential authentication (e.g., passwords, identification number, PINs, etc.). Standalone credential authentication is not considered secure, granted that it can be bypassed using brute forcing and fuzzing techniques. Additionally, phishing, spear-phishing, vishing, etc., and other social engineering campaigns have also managed to exfiltrate sensitive authentication data from operators potentially jeopardizing the secure operation of wind-based DERs. }% or chemical combinatorial attacks.

The electrification of the transportation sector is a major thrust towards grid decarbonization. As a result, a precipitous growth has been observed in the number of EVs across the world. %Among their advantages (e.g., greener form of transportation),
EV batteries can be used as backup power supplies during unexpected and instant power demand. To support such ancillary backup services, as well as to charge their battery resources, EVs connect to the electric grid via charging stations. According to \cite{raju2019}, EV charging stations (EVCS) % perform the following tasks. 
first, authenticate the EVs, and then they either charge them or connect them to the main grid to be used as ad-hoc energy storage. Authorization can be done through radio frequency identification (RFID), Bluetooth, or Wi-Fi technologies. However, RFID systems have been proven to be insecure and are vulnerable to malicious attacks such as eavesdropping and active interference \cite{aikaterini2008}. RFID readers and tags operate in noisy environments, which decreases security and can consequently compromise the EVCS. Furthermore, authorization done using wireless networks (Wi-Fi or Bluetooth) can be exploited by adversaries as well. Multiple design flaws, security weaknesses, and practical attack scenarios have been reported regarding Bluetooth technologies, enabling adversaries to pair to devices and impersonate legitimate users \cite{haataja2010two}. Apart from the authorization and wireless network configuration security oversights, % a plethora of threats targeting EVs is reported 
in \cite{mustafa2013smart} the authors present the vulnerable components of EVs that could be exploited by adversaries (e.g., remote attacks) compromising grid operation.

\textcolor{black}{Apart from the discussed vulnerabilities concerning generation-type DERs, critical vulnerabilities in DER controllable loads also exist. The IoT is a network of connected devices including smart appliances (e.g., smart thermostats, air-conditioning, heat pumps, EVCS points, etc.), enabling the exchange of information between devices and users leveraging wired and wireless connections \cite{xenofontos2021consumer, lakshminarayana2022load}. These IoT-enabled devices are connected to the load-end of DERs and their operation is typically orchestrated remotely by their end-users. According to \cite{xenofontos2021consumer, makhdoom2019}, a large number of IoT devices (using remote internet access mechanisms) still lack concrete security mechanisms and, as a result, can be vulnerable to cyberattacks. Researchers have demonstrated that attackers can manipulate IoT-connected devices in low-inertia EPS (e.g., relying on renewable generation) and other exogenous conditions to cause unsafe grid operation. That is, in \cite{lakshminarayana2022load, ospina2020feasibility}, the authors investigate the impact on grid stability of load altering attacks -- if multiple IoT-connected high-wattage devices are compromised simultaneously -- during the COVID-19 lockdown period. This adversarial behavior is enabled by the absence of overarching defense mechanisms, such as anti-viruses, and the sporadic provision of software/firmware updates and security patches which further weaken the device security posture. As a result, the vulnerabilities of such IoT-connected devices can be ported to the utility grid, given their interconnection to DERs. }

The authors in \cite{benjamin2019} report that another compelling reason undermining IoT -- and DER-controllable load -- security is that during the design cycles, function and lowering manufacturing costs were prioritized above firmware security.  As a result, end-users are presented with feature-full devices with multiple backdoors that adversaries can exploit; paving the way for malware developers \cite{benjamin2019}. Smart thermostats, for example, represent a prime example of how the consequences of a compromised IoT device could propagate, posing threats to the power grid. Smart thermostats regulate the temperature of residential or commercial facilities by ``cleverly'' managing heat sources and air conditioners. Thermostats can learn their users' patterns and adapt their operation accordingly. The cybersecurity of such devices has been shown to be lacking since many are shipped from manufacturers using default configurations and predefined credentials \cite{xenofontos2021consumer}. Furthermore, the end-users of such devices might be unaware of the essential steps %to configure their devices and 
secure their devices against cyberattacks. Consequently, such devices are left vulnerable to attacks that could simultaneously trigger on/off particular loads (e.g., heaters) deliberately affecting the instantaneous power demand of power grids, leading to blackouts or brownouts \cite{soltan2018blackiot}.

\textcolor{black}{Similarly to IoT-based systems which leverage ICTs to exchange information, recently the design of solar farms has also incorporated ``smart'' features. In large solar deployments, ICTs are used for the remote and real-time management of the generated power used for economic dispatch and demand-response schemes. Centralized storage solutions are also used for the aggregation of historical and generation data, exposing solar farms to threats (e.g., single-point-of-failure paragon). The remote control, communication, and data aggregation features of grid resources can be exploited as attack entry points. This was the case with the cyberattack targeting VESTAS offshore wind turbine manufacturer, which led to the ex-filtration of sensitive data and ``\emph{forced VESTAS to shut down IT systems across multiple business units and locations to contain the issue}'' \cite{vestas}. Although, according to the Danish wind manufacturer, customers were not affected, the socio-economic and reputation blow to the company was acute. }

\textcolor{black}{In residential solar deployments, communication technologies such as Bluetooth are used regardless of the limited cybersecurity mechanisms, which can result in security breaches compromising user privacy \cite{cardenas2020assessing}. Aside from the attacks on IoT controllable loads and generation assets that stem from inherent implementation or architectural vulnerabilities, the possibility of supply-chain-based attacks should also be considered. Security oversight in manufacturing facilities can be exploited to develop supply-chain attacks, in which adversaries are granted apriori control over thousands of DER-connected assets \cite{supply}. The possibility of supply-chain compromises along with the catastrophic consequences of such events was demonstrated during the colonial pipeline incident in 2021 \cite{colonial}. }

\subsection{DER Device Level Attacks} \label{sec:device_attack}
Attacks targeting the DER device level exploit implementation, architectural, and other security design oversights (e.g., communication, internal storage, remote update or control functionality, etc.) to maximize their impact and compromise EPS operations. In the case of inverters, for example, after successfully gaining access to the DER assets, adversaries can launch cyberattacks including, \emph{i)} DoS attacks compromising the inverter's availability, \emph{ii)} data alteration attacks where the exchanged data between DERs and utilities are maliciously modified, and \emph{iii)} command injection attacks where termination commands or malicious controls are forwarded to the inverters \cite{kang2015investigating, ye2021review}. During DoS attacks, the communication bandwidth of the inverter can be flooded leading to disruptions, i.e., the process control flow of the device is suspended \cite{carter2017cyber}. Inverter operation is dynamically regulated to enhance power generation efficiency and support ancillary services when requested by grid operators. However, adversaries could delay or prevent the transmission of critical data (e.g., sensor measurements, control commands, etc.) to certain DERs inhibiting their operation \cite{carter2017cyber}.

\textcolor{black}{MitM attacks can also be leveraged to compromise grid inverters. In such scenarios, attackers can gain device-level information by eavesdropping on data traffic between inverters and the utility grid. Attackers can access system information by decoding MMS real-time data packets \cite{teymouri2018cyber}. %MMS is used to transmit real-time data between connected devices in a network. 
Given that most inverter models use MMS to send and receive data, such attacks could have a considerable impact on grid stability. Attackers could compromise the DER operation by modifying the inverter configuration using undesirable parameters, e.g., tampering with the reactive power references \cite{ferc_react}. %of inverter-based DERs  
Brute force attacks can also be leveraged to compromise DER assets. In \cite{asal2018}, researchers demonstrate that adversaries can brute force weak PIN sequences and get access to WTCPs. Similarly to the inverter command injection attack, adversaries can send malicious commands to wind turbines, modifying setpoints or control objectives, causing unexpected operations \cite{STAGGS20173}. }

\textcolor{black}{In many cases, DER assets are owned by end-users, %, e.g., rooftop solar, EVs, battery storage, etc. 
%In such scenarios, DERs can be connected 
interfaced to user networks, and exhibiting similar operation to common IoT devices. As such, DERs could be remotely operated and controlled by consumers or other third-party applications. IoT devices have been exploited on a massive scale in the past, as was the case with the Mirai botnet attack \cite{xenofontos2021consumer}. The Mirai malware, after its propagation to multiple vulnerable IoT devices, grants adversaries full control over the compromised devices. Groups of compromised devices, i.e., botnets, can then be collectively exploited to cause distributed DoS (DDoS) attacks \cite{gopal2018}. However, in the case of DERs, %which have power generation and ancillary service capabilities (e.g., reactive power compensation), 
such attacks could severely affect grid stability. Apart from botnet attacks, other IoT-sourced vulnerabilities could be also abused.% compromising DER operation. 
Threat actors can perform replay attacks by replicating legitimate commands transmitted from DER asset owners. RFID authentication and Bluetooth or Zigbee communication packets can be eavesdropped and replayed, enabling unauthorized access to DER devices \cite{karim2020}. The security of these IoT-connected DER devices is relied upon the user competence in safeguarding their assets. If IoT network security is overlooked, adversaries could capture exchanged data, obtain session or encryption keys, access the devices, and issue commands to render DERs unreachable (DoS) or instruct anomalous operations \cite{karim2020}. }

\vspace{-3mm}
\subsection{DER Device Level Impacts} \label{sec:device_impact}

Adversaries, after gaining initial access to DER devices, can follow different approaches to deploy their attacks \cite{MITRE_ICS}. If maximizing the \emph{immediate} system-wide impact is the objective, overvoltage or undervoltage conditions, frequency fluctuations, false trippings, and disconnection mechanisms could be targeted. Power systems have built-in mechanisms to automatically detect and isolate such high-impact events to limit their consequences. Different detection approaches (e.g., physics-based, data-driven, or a combination of the two) could be used to overcome such conditions \cite{teymouri2018cyber, zografopoulos2021security, konstantinou2021resilient}. 

\textcolor{black}{On the other hand, in the advanced persistence threat (APT) case, threat actors might prioritize system persistence, breach of privacy, and long-term system degradation instead of immediate impact, and opt for more sophisticated and stealthy attacks \cite{APT}. For instance, attackers could exploit firmware vulnerabilities and achieve remote access on DERs through public-facing applications or the supported remote services \cite{MITRE_ICS}. While in control of the DER device, attackers can stealthily perform minute modifications to system parameters or coordinate attacks in ways that will not affect the net system behavior deceiving detection mechanisms \cite{rath2021stealthy, sahoo2020ondetection}. Such stealthy attacks might cause unsafe, unstable, or uneconomic operation of IBRs. Adversaries could also exfiltrate sensitive user information (e.g., credentials, passwords, etc.), since DER devices might be connected in user-owned home networks. Attackers may also be learning the operational patterns of DERs, that is, aggregating enough system information to identify the temporal and spatial conditions which, if satisfied, can maximize the impact of attacks on the grid~\cite{rath2021stealthy, Rath2022behind}. }%\hl{we could include the updated work on this topic if PESGM paper is accepted}

The rapid adoption of EVs (from 3 million in 2017 to 125 million by 2030 \cite{bunsen2018global}) makes them prominent targets for attackers aiming to gain access to EVs or EVCS. Malware can be deployed and propagated throughout the whole EV infrastructure, compromising the charge controllers, demand-response schemes, charging limits, grid power quality, and crippling the power and transportation sectors. %\cite{park2019potential}. 
According to \cite{acharya2020public}, compromised EV supply equipment (EVSE) servers can prevent EVCS sessions by denying authentication, or reproducing incorrect information about charging station data (e.g., price, online station status, etc.). Furthermore, EVCS being high-wattage assets, if maliciously manipulated in addition to power-related consequences (e.g. increased load demand, power quality issues, etc.), can also inflict considerable financial losses on electric power utilities \cite{acharya2021cyber}. Similarly, sensitive user information (e.g., identity, location, payment information, etc.) could be leaked during potential attacks on the EV and charging infrastructures \cite{8994200}. The absence of attack impact assessment methodologies for EV networks is highlighted by the authors in \cite{8790593}. Their work attempts to identify potential attack and failure scenarios while demonstrating the consequences of such events on the corresponding EV system security. \looseness=-1

Researchers have demonstrated that there are multiple paths to compromise the security of wind-integrated DER assets. In \cite{asal2018}, access to WTCP is granted by launching a brute force dictionary attack on the WTCP device. Similarly, in \cite{STAGGS20173}, by compromising network devices and wind process automation controllers, malicious requests can be sent to turbines, preventing nominal operation and potentially causing damage to critical electrical and mechanical components. The impacts of such attacks can lead to fires, explosions, jeopardize personnel assigned to resolve such issues, and the safety of surrounding communities %. Such impacts are highlighted by a plethora of high-profile wind incidents reported in the literature 
\cite{windAccidents3, windAccidents4}. The aforementioned high-profile wind incidents might have not been caused by cyberattacks, but the probability of exploiting wind turbine vulnerabilities remotely exists. Furthermore, the impact that attacks on wind turbines could have on grid reliability has been investigated in the literature \cite{yan2011cyber, zhang2016power}. \looseness=-1

\textcolor{black}{Malware, such as the Mirai and its derivatives, or other supply-chain exploits (e.g., hardware trojans, vulnerabilities in commercial and open source software) could be leveraged to control the operation of IoT-connected DERs leading to intermittent operation and DDoS attacks \cite{ahmed2019}. As a result, the botnet of the compromised DER devices can be maliciously operated as a distributed load or generation. Grid instability can thus occur when DERs and other remotely controlled high-wattage devices (i.e., smart thermostats, EVs, EVSE, Heating, Ventilation, and Air Conditioning -- HVACs, etc.) are simultaneously switched on or off, severely affecting the electricity power demand \cite{huang2019not}. In large-scale demand-side attack scenarios, grid operators will be forced to perform load shedding to sustain critical loads \cite{ospina2020feasibility}. }
%A summary of potential vulnerabilities and the attacks impacts of demand-side compomises targeting DER devices and aggregators is depicted in Fig. \ref{fig:LAA_attack}. 
\textcolor{black}{Consumers might experience brownouts and service interruptions, while the impacts of load-altering attacks on the local power supply will also affect demand-response schemes \cite{dang2019demand}. In addition to grid impacts, infected devices could also propagate malware (in a worm-like fashion) to the rest of the devices residing within the same network (e.g., switches, personal computers, mobile phones), thus, jeopardizing sensitive user information \cite{ronen2017iot}. }

\subsection{DER Device Level Mitigations} \label{sec:device_mitigation}
\begin{figure*}[t]
\centering
\includegraphics[width=0.85\linewidth]{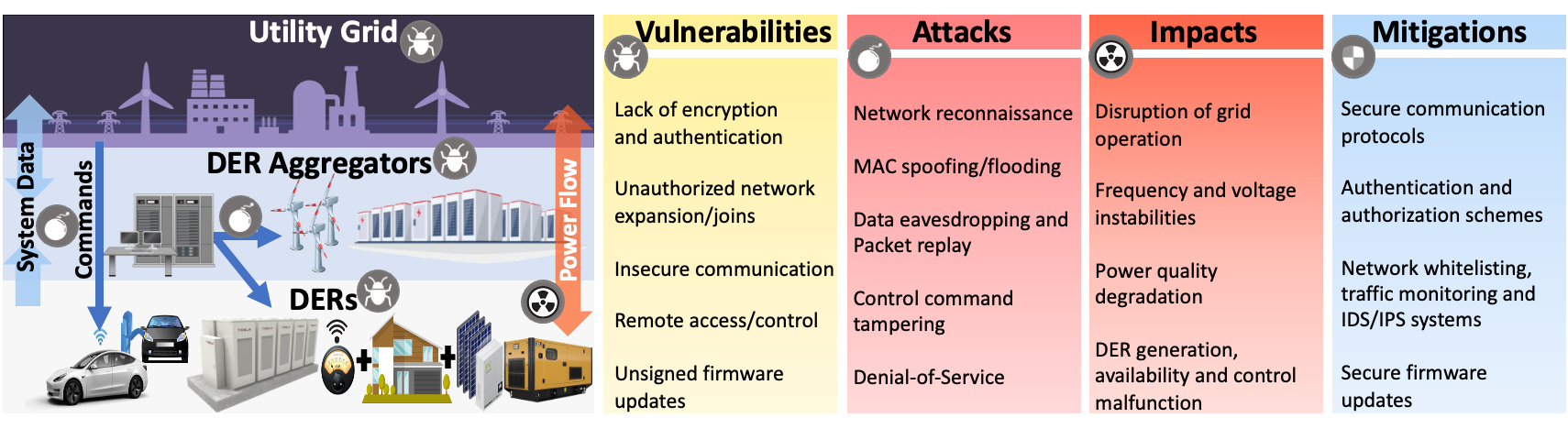}
\caption{Overview of DER-integrated electric grid that illustrates how the layered architecture expands the threat surface.}
\label{fig:LAA_attack}
\vspace{-3mm}
\end{figure*}

\textcolor{black}{To combat threats targeting DER devices, a multitude of mitigation strategies have been proposed. With respect to inverters, the authors in \cite{qi2016cybersecurity} suggest that an additional power electronic interface (energy buffer) should be developed and placed at different locations within the distribution grid (mainly at points where the device is connected to the grid). This energy buffer interface can help avoid unintentional islandings, which could be triggered due to the conflicting anti-islanding detection and low/high voltage ride-through functions that inverters support. The micro-architectural hardware components of inverter controllers have also been utilized for the detection of malicious events within the DER asset's process control loop \cite{kuruvila2021hardware}. In \cite{zografopoulos2021detection}, the inherent physical properties of DER-integrated grid systems are used to identify potential actuator or sensor modifications which could lead to abnormal system operation. Furthermore, a multitude of model- and data-driven anomaly detection methods have also been proposed to identify compromised DER assets \cite{li2016active}. % ,  li2019detection}. 
Once intrusions have been efficiently detected, the compromised assets are isolated and the available grid resources are utilized to achieve resilience and maintain the power supply-demand balance  equilibrium \cite{bidram2019resilient, 9127110}. % \cite{7307203, abhinav2017synchrony, bidram2019resilient, 9127110}
}

\textcolor{black}{The prevention of adversaries from launching DoS attacks on EVSE can be achieved via the use of hardened operating systems  \cite{ricardo2019}. Such systems can be leveraged to deploy and keep the EVSE with up-to-date firmware. Furthermore, roll-back firmware update functionality is essential, enabling the use of previously working firmware, in cases where the updated firmware versions are proven unreliable or have been maliciously modified. Different methodologies have also been proposed for the mitigation of the harmonic interference of EV-related cyberattacks. For instance, the authors in \cite{7438898} and \cite{6138922}, propose pulse-width-modulation and a high-frequency resonance scheme, respectively, that can mitigate the power quality degradation introduced by EVs and bidirectional grid-tied converters (e.g., battery storage). On the other hand, a power management framework leveraging renewable resources and the EV infrastructure itself is used in \cite{7579545}, in order to alleviate power quality challenges encountered in unbalanced distribution networks. The importance of cyber insurance against cyberattacks on EVCS has also been highlighted by researchers. In \cite{acharya2021cyber}, the authors demonstrate that defense mechanisms and cyber insurance policies can effectively reduce the financial impacts of cyberattacks and curtail insurance premiums for the entities that manage EVCS.}

Towards mitigating brute force attacks targeting weak credentials used by WTCP, \cite{ricardo2019} recommends the usage of password management systems. Furthermore, monitoring user behavior (e.g., failed login attempts), network segregation, and role-based access control rules are encouraged. % to restrict adversarial access to wind DER assets. 
To mitigate stealthy attackers who aim to gain access to wind turbine controllers exploiting their remote firmware update capabilities (e.g., over-the-air updates), in \cite{lee2017blockchain} and \cite{falas2019hardware} different methodologies are proposed to verify firmware trustworthiness. In \cite{lee2017blockchain}, a secure firmware update scheme is introduced, where devices leverage the blockchain to check for new firmware versions and validate their integrity before downloading and installing the firmware images. On the other hand, in \cite{falas2019hardware}, the authors leverage a hardware architecture to validate firmware authenticity. Cryptographic modules %, in addition to the embedded device's physical characteristics, 
are utilized to guarantee the integrity and authenticity of firmware packages. \looseness=-1
% impinging malicious firmware deployment attacks.

\textcolor{black}{When viewing DERs from the IoT perspective, most of the security implications and attack mitigation strategies applicable for IoT end-devices could be exercised. In \cite{savola2016}, the authors suggest the use of a personal security application (PSA) as a countermeasure. % for IoT-related attacks. 
PSA resembles a combination of security features, such as access control mechanisms, malware detection, network traffic, and resource utilization monitoring, etc., to strengthen the security posture of IoT devices and networks. Blockchain technology has also been mobilized to safeguard the security of IoT networks, singling out malicious nodes from benign ones \cite{9540839}. For example, a blockchain detection methodology is presented in \cite{ahmed2019} to protect IoT nodes from the Mirai botnet.} 

\textcolor{black}{On the other hand mitigating supply-chain compromises, hardware- or software-based can be challenging. State-funded initiatives, such as the CHIPS program in the USA are essential to establish ``a secure and resilient semiconductor supply-chain that adheres to standards and guidelines on information security, data tracking, verification, and promotes the further development and adoption of such standards" \cite{CHIPS}. Furthermore, adhering to recommendations and practices issued by cybersecurity organizations, such as the National Security Agency (NSA), the Cybersecurity and Infrastructure Security Agency (CISA), etc. is crucial to mitigate software supply-chain liabilities \cite{CISA_supply}.} In \cite{xenofontos2021consumer}, %a thorough review of IoT attacks and their corresponding mitigations are presented. 
best practices, mitigation techniques, and security policies are enumerated, viewed from the IoT device, infrastructure, communications, and services standpoints %The authors discuss open challenges regarding the security of future IoT systems, 
which closely match the obstacles that DER infrastructure will have to overcome.% (e.g., vendor locking, computationally limited resources, cost, data security, etc.).  

%%%%%%%%%%%%%%%%%%%%%%%%%%%%%%%%%%%%%%%%%%%%%%%%%%%%%%%%%%%%%%%%%%%%%%%%%%%%%%%%
%%%%%%%%%%%%%%%%%%%%%%%%%%%%%%%%%%%%%%%%%%%%%%%%%%%%%%%%%%%%%%%%%%%%%%%%%%%%%%%%
%%%%%%%%%%%%%%%%%%%%%%%%%%%%%%%%%%%%%%%%%%%%%%%%%%%%%%%%%%%%%%%%%%%%%%%%%%%%%%%%
%%%%%%%%%%%%%%%%%%%%%%%%%%%%%%%%%%%%%%%%%%%%%%%%%%%%%%%%%%%%%%%%%%%%%%%%%%%%%%%%
%\section{DER Cybersecurity Evaluation and Remarks} 
\section{DER Cybersecurity Concluding Remarks} 
\label{s:conclusion}
\vspace{-1mm}

\begin{table*}
\centering
\small
\caption{Cyberattack mitigations and design best practices.}
\vspace{-2mm}
\label{table:mitigations}

\renewcommand{\tabularxcolumn}[1]{m{#1}}

\begin{tabularx}{\textwidth} { 
  || >{\hsize=1.90\hsize\linewidth=\hsize\raggedright\arraybackslash}X 
  | >{\hsize=0.70\hsize\linewidth=\hsize\centering\arraybackslash}X
  | >{\hsize=0.40\hsize\linewidth=\hsize\centering\arraybackslash}X || }
  
\hline \hline
{Mitigations} & {Attacks} & Reference \\ 

\hline
{Connection auditing using authentication, authorization, and accounting servers, role-based access control, MAC address white-listing, %for network switches  
unused port hardening} & {MAC spoofing/flooding, DoS, MitM, SYN flooding}  &  {\cite{sundararajan2018survey}, \cite{carter2017cyber}}\\ 
\hline

{IP header (Sequence number) inclusion in TCP, cryptographically enhanced address resolution protocols (ARP), secure key distribution schemes} & {Packet replay, DoS, MitM, SYN flooding}                                    &  {\cite{qi2016cybersecurity}, \cite{carter2017cyber}}\\ 
\hline

{Firewalls, intrusion detection/prevention systems (ID/IPS), traceback and push-back services, cryptographic hashing and stack tweaking}  & {MAC flooding, DoS, MitM, SYN flooding}                  &  {\cite{qi2016cybersecurity}, \cite{lai2017cyber}, \cite{sundararajan2018survey}} \\ 
\hline

{Password management systems, access restriction after multiple failed log-in attempts, hardened operating system kernels, roll-back firmware updates} & {Brute force attacks, DoS, Packet replay, Eavesdropping} & {\cite{ricardo2019}}\\ 
\hline

{Personal security and privacy practices (e.g., security updates, password managers, encryption, ephemeral keys, etc.)}  & {Packet replay, Eavesdropping} & {\cite{savola2016}}\\ 

\hline \hline
\end{tabularx}

\end{table*}

\textcolor{black}{Improving EPS cybersecurity and resilience is of critical importance and DERs are expected to contribute toward this effort. However, the sophistication of cyberattacks targeting EPS indicates that multifaceted security approaches should be considered. In Fig. \ref{fig:LAA_attack} we present potential vulnerabilities at the different levels of the grid architecture, and the impacts of attacks targeting DER devices, aggregators, and utilities. 
Detection, protection, and mitigation schemes should factor in the inherent vulnerabilities in both DER stacks (cyber and physical) and on every level, from DER assets to system operators. Therefore, security evaluation metrics, which consider how DERs contribute to EPS, are crucial, given the potential risks and system-wide impacts of cyberattacks. }

\subsection{Cybersecurity Metrics for DER-integrated Systems} \label{s:metrics}
The risks associated with compromises involving DER assets require substantial cybersecurity investments from utilities and aggregators. To justify such investments, metrics are essential for assessing the cybersecurity posture of the involved stakeholders, and the efficiency of the implemented cybersecurity methodologies in curtailing potential risk. \textcolor{black}{In \cite{paul2021vulnerability}, diverse analytical metrics are discussed to assess the resilience of cyber-physical systems (CPS). Similarly, in \cite{haque2018cyber} and \cite{jacobs2018measurement} cybersecurity and resilience metrics are developed to analyze industrial control systems. Different approaches have been followed to provide qualitative and quantitative metrics to measure the cybersecurity and resilience of power systems \cite{venkataramanan2019cp, venkataramanan2020tram, johnson2020assessing}. For instance, NIST has proposed a framework that consists of five essential functions to overcome adverse events, namely identify, protect, detect, respond, and recover \cite{sedgewick2014framework}.} For each of these five functions, performance metrics are then employed to assess the ICS security posture using insights from real-world incidents. However, such frameworks fail to measure security holistically since only specific system sections are investigated. \looseness=-1

\textcolor{black}{To address this issue, in \cite{haque2019cyber}, the authors extend the \emph{R4} resilience framework \cite{tierney2007conceptualizing} by introducing a quantitative metrics hierarchy under four main domains, i.e., system robustness, redundancy, resourcefulness, and rapidity. CPS cybersecurity is then evaluated by decomposing each of the four aforementioned domains into subcategories and deriving scores based on asset criticality, interconnectivity, system network topology, and underlying physical processes. A similar path is followed in \cite{ospina2022cpes} where a quantitative security metric is proposed that factors the interaction between cyber and physical layers. Then, optimal decisions are made by integrating this security metric in cyber-constrained AC power flow studies.\looseness=-1}

\textcolor{black}{EPRI highlighting the challenging task of quantifying security in diverse DER architectures has presented a data-driven cybersecurity metric methodology \cite{EPRImetrics}. The proposed metrics framework combines real-world IT and OT data aggregated from the system-under-test. Sixty security metrics, including mean time to discovery, mean time to recovery, threat awareness, endpoint protection scores, etc., are then combined to assess and quantify the cybersecurity status of the system. The sixty metrics are categorized under three core categories (i.e., operational, tactical, and strategic) depending on the operational constraints and security requirements as identified by each stakeholder (e.g., utility, aggregators, etc). EPRI has open-sourced its cybersecurity metrics calculation software, \emph{OpenMetCalc}, which allows users to load their system-specific data and compute their system's performance with respect to these sixty metrics \cite{OpenMetCalc}. Based on these scores, executive decisions can be made, and resources can be prioritized to reduce and/or mitigate potential risks. Although this cybersecurity metric endeavor led by EPRI is a step in the right direction, it is still an ongoing research topic that requires active participation from the energy sector.}

\vspace{-2mm}

\subsection{Future Challenges} \label{s:conclusionsAndFuture}
\textcolor{black}{This work explores the cybersecurity posture of DERs as an essential building block 
%given their rapid penetration and significance 
for future resilient EPS. \textcolor{black}{We investigate threats and present an overview of attacks % targeting DER assets. 
without focusing on specific DER types, e.g., rooftop solar, BESS, or wind turbines. DER vulnerabilities are examined from the protocol and device levels, which are pertinent regardless of the DER type. We furnish a consolidated review of attacks and their potential impacts. % on DER operations. 
% if such vulnerabilities are maliciously exploited. 
Mitigation methodologies and design best practices are also discussed, and in Table \ref{table:mitigations}, we compile a summary of mitigation schemes against different attack types.\looseness=-1}}

\textcolor{black}{Even though the proposed strategies could reduce the DER threat surface, they cannot be considered as ``silver bullet solutions''. % which can address every application. 
This fact is partly attributed to the DERs' distributed mode of operation, which can be utility-, aggregator-, or prosumer- owned. Especially for the latter case, user negligence in the security configuration of their DER assets, could give rise to compromises. However, the risk and magnitude of such disturbances depend on how many DERs could be attacked simultaneously. Security standards and policies -- such as IEEE 1547-2020 \cite{IEEE1547}, NISTIR 7628 \cite{NISTIR_7628}, NIST SP800-82 \cite{NISTSP800}, CA Rule 21, Hawaii Rule 14 \cite{Rule_14H}, and IEEE 1815.1 \cite{IEEE1815} -- should be enforced ensuring the innocuous DER operation.} %even in untrusted environments.}% (e.g., user-owned assets).

\textcolor{black}{The strong coupling between electricity markets, demand response schemes, and DERs can also incentivize the exploitation of DERs for financial benefits. False data injection attacks manipulating measurement points at the distribution level have been demonstrated to be capable of deceiving state estimators and influencing economic dispatch mechanisms \cite{liu2019financially}. Load forecasts can thus be biased, affecting the marginal prices in electricity markets, to benefit corrupt distribution utility operators and DER aggregators. To thwart such attacks, resilient state estimation algorithms \cite{konstantinou2021resilient}, and incentive reduction policies should be implemented \cite{liu2019financially}.}

\textcolor{black}{ Security challenges will still exist, regardless of the preventive and preemptive methodologies that we propose. % and the security of future DER devices that is expected to improve (both at the device and protocol levels). 
Developing universal risk management schemes can be a perplexing or infeasible task due to the distributed, ad-hoc (e.g., EVs, battery storage) and stochastic (e.g., solar inverters) nature of DERs. However, comprehensive system modeling using digital twin systems and data-driven approaches can be leveraged to forecast grid behavior and predict cyberattack impacts. Additionally, risk assessment and cyber-physical risk metrics can be used to evaluate and prioritize mitigation decisions. High-fidelity information, derived from system models, can assist risk metric estimations, which can prove useful when designing response strategies and self-healing schemes \cite{zografopoulos2022mitigation}.}

The influx of DER devices and their projected numbers underscore the need for comprehensive security practices. To harness DER advantages and withstand their underlying security impediments, the combined knowledge of security engineers, the industry, and academia is essential. User vigilance is also crucial to impede malicious behavior attempting to achieve foothold on user-owned DERs. % and even post-compromise to restrict attack propagation. 
Security standardization and policy-making procedures can strengthen the cybersecurity posture of DERs and prevent vulnerabilities from materializing into threats. \textcolor{black}{However, if the discussed practices fail to prevent or detect attacks, risk metrics, detailed system modeling, and mitigation plans can orchestrate resources to inhibit or overcome undesirable grid conditions and enhance EPS resilience.}

\section*{Acknowledgement}
This publication is based upon work supported by King Abdullah University of Science and Technology (KAUST) under Award No. ORFS-CRG11-2022-5021.

\vspace{-2mm}
\bibliographystyle{IEEEtran}
\bibliography{biblio_short}

\end{document}